\documentclass[onecolumn]{aastex701}

\newcommand{\FeII}{Fe~{\sc ii}}

\newcommand{\kms}{\hbox{km~s$^{-1}$}}
\newcommand{\cmsq}{\hbox{cm$^{-2}$}}

\newcommand{\flux}{\hbox{erg~cm$^{-2}$~s$^{-1}$}}

\newcommand{\lumin}{\hbox{erg~s$^{-1}$}}

\newcommand{\nh}{\hbox{${N}_{\rm H}$}}

\newcommand{\h}{$h_{70}^{-1}$}

\newcommand{\be}{\begin{equation}}
\newcommand{\ee}{\end{equation}}
\newcommand{\ba}{\begin{eqnarray}}
\newcommand{\ea}{\end{eqnarray}}

\newcommand{\nustar}{\emph{NuSTAR}}

\newcommand{\swift}{\emph{Swift}}

\newcommand{\simgt}{\lower 2pt \hbox{$\, \buildrel {\scriptstyle >}\over {\scriptstyle\sim}\,$}}
\newcommand{\simlt}{\lower 2pt \hbox{$\, \buildrel {\scriptstyle <}\over {\scriptstyle\sim}\,$}}
\newcommand{\ls}{\lower 2pt \hbox{$\;\scriptscriptstyle \buildrel<\over\sim\;$}}
\newcommand{\gs}{\lower 2pt \hbox{$\;\scriptscriptstyle \buildrel>\over\sim\;$}}

\begin{document}

\title{A Decade-Long Increasing Mid-Infrared Luminosity in Galaxy NGC6447: a Turning-On Candidate of Active Galactic Nucleus}

\author[0000-0001-9203-2808]{Xinyu Dai}
\affiliation{Homer L.\ Dodge Department of Physics and Astronomy,
University of Oklahoma, Norman, OK 73019, USA}
\email{xdai@ou.edu}

\author[0009-0007-3655-1444]{Nate Adams}
\affiliation{Homer L.\ Dodge Department of Physics and Astronomy,
University of Oklahoma, Norman, OK 73019, USA}
\email{}

\author[0009-0002-4279-9925]{Natalie Kovacevic}
\affiliation{Homer L.\ Dodge Department of Physics and Astronomy,
University of Oklahoma, Norman, OK 73019, USA}
\email{}

\author[0009-0006-1659-7472]{Kaitlyn Parrinello}
\affiliation{Homer L.\ Dodge Department of Physics and Astronomy,
University of Oklahoma, Norman, OK 73019, USA}
\email{}

\author[0000-0003-3726-5611]{Marko Mi\'ci\'c}
\affil{Homer L.\ Dodge Department of Physics and Astronomy,
University of Oklahoma, Norman, OK 73019, USA}
\email{}

\author[0000-0002-7720-3418]{Heechan Yuk}
\affil{Homer L.\ Dodge Department of Physics and Astronomy,
University of Oklahoma, Norman, OK 73019, USA}
\email{}

\author[0009-0002-7796-3406]{Zijun Gao}
\affiliation{Homer L.\ Dodge Department of Physics and Astronomy,
University of Oklahoma, Norman, OK 73019, USA}
\email{}

\author[0009-0005-9187-4022]{Lorelei Starling}
\affiliation{Homer L.\ Dodge Department of Physics and Astronomy,
University of Oklahoma, Norman, OK 73019, USA}
\email{}

\author[0000-0001-8973-5051]{Francesco Shankar}
\affiliation{School of Physics and Astronomy, University of Southampton, Highfield, Southampton 1BJ, UK}
\email{}

\begin{abstract}
It is widely expected that the obscured accretion stage can be the initial turning-on stage of active galactic nuclei from quiescent galaxies.
We present mid-infrared light curves of NGC 6447 in 3.5$\mu$m and 4.6 $\mu$m bands observed by WISE/NEOWISE, which show an almost monotonic increasing trend of 1.2 mag over 14 years.  
The optical light curve from ASAS-SN during the same period is consistent with a constant showing no variability.
The mid-infrared color evolution shows that the galaxy transitioned into an active galactic nucleus (AGN) in 2018.
The SPHEREx spectrum reveals an increasing continuum resembling warm to hot dust emission from an AGN.
NuSTAR detected an X-ray source with a 2--30~keV luminosity of $8.4\times10^{41}$~\lumin\ at the lower boundary of AGN X-ray emission range, and a factor of $>7$ variability in one year compared to the Swift upper limit.
NGC 6447 was classified as a quiescent galaxy in the literature.  The multi-wavelength timing and spectral properties of NGC 6447 are consistent with the expected AGN turning on event,
where the obscuring material around the AGN central engine is gradually dispersed, revealing the central engine.
This example shows that long-term infrared variability can be a powerful tool to find similar sources.
Based on the sample selection statistics, we estimate the duration of the episodes of AGN accretion (duty cycle) signified by the turning-on event as $10^4$--$10^6$~yr.

\end{abstract}

\keywords{\uat{Active galactic nuclei}{16} --- \uat{Accretion}{14} ---\uat{Galaxy evolution}{594} --- \uat{Supermassive black holes}{1663} --- \uat{Time domain astronomy}{2109}}

\section{Introduction} 
Supermassive black holes (SMBHs) are believed to reside in the centers of the most massive galaxies, yet only a minority are observed while actively accreting the surrounding material, producing a powerful source known as the active galactic nucleus (AGN). Multiwavelength studies demonstrate that the fraction of galaxies hosting nuclear activity varies depending on the observational tracer, with X-ray surveys, infrared diagnostics, radio emission, optical spectroscopic classifications, and variability typically yielding AGN occupation fractions ranging from a few to a few tens of percents   \citep{2003MNRAS.346.1055K,2009MNRAS.398.1165G,2010ApJ...723.1447H,2012ApJ...746...90A,2013MNRAS.429.1827P, 2020AJ....159...69M, 2022ApJ...930..110Y}. These results imply that AGN activity operates in an episodic manner, with SMBHs switching between active and quiescent states over cosmic time \citep{2014ApJ...782....9H, 2015MNRAS.451.2517S}. Although the global demographics and energetic impact of AGN have been well investigated, the physical mechanisms that trigger AGN activity and the observational signatures of their earliest stages are poorly understood and understudied. The particularly important stage in which the SMBH transitions from quiescent to active state, the so-called AGN turn-on phase, is expected to be short-lived and heavily obscured, making it notoriously challenging to identify. Capturing an AGN during this phase is essential for understanding how gas is delivered to the nucleus, how obscuration evolves, and how feedback begins to operate.

Theoretical models and numerical simulations suggest multiple pathways for triggering AGN activity. Major mergers of gas-rich galaxies can efficiently drive large quantities of gas toward the nuclear region through gravitational torques, potentially igniting luminous quasar activity \citep{1991ApJ...370L..65B,2006ApJS..163....1H,2008ApJS..175..356H,2023MNRAS.522.1736P,Breiding_2024}. However, in the local universe and at moderate AGN luminosities, secular processes are thought to play a dominant role. Bars, spiral density waves, minor interactions, and disk instabilities can transport gas inward over longer timescales, fueling lower-luminosity AGN \citep{1990Natur.345..679S,1993A&A...271..391C,1997ApJ...486..750Y,1998ApJ...509...85B,2011ApJ...743L..13C,2012ApJ...748L...7V,2015MNRAS.447.2123C}. In both scenarios, AGN fueling is expected to vary with time, causing accretion rates to fluctuate on timescales ranging from years to millions of years \citep{2011ApJ...741L..33B,2011ApJ...737...26N,2013MNRAS.434..606G}. Therefore, AGN activity likely proceeds via multiple short episodes rather than a single continuous phase, complicating the identification of the initial onset of activity, particularly if early accretion is dust-obscured.

Several observational classes have been proposed as candidates for young AGN. The most notable example is the population of red quasars, which exhibit a reddened optical and near-infrared continuum, often interpreted as an early AGN evolutionary phase emerging from a dust-enshrouded stage following a merger-driven inflow, before feedback clears the surrounding material. \citep{1995Natur.375..469W,2007ApJ...667..673G,2012ApJ...757...51G}. Similarly, broad absorption line (BAL) quasars have been suggested to represent a young or transitional AGN phase associated with strong winds that expel gas and dust from the nuclear region, gradually revealing the central engine \citep{1993ApJ...413...95V, 2007ApJ...662L..59F, 2009ApJ...698.1095U, 2012ApJ...757..180D}. Other proposed early-stage AGN candidates include infrared-luminous, optically faint galaxies, in which the bolometric output is dominated by dust-reprocessed emission \citep{2004ApJS..154..166L,2012ApJ...748..142D}. Nevertheless, most known young AGN candidates are identified during luminous or already relatively well-developed AGN phases. As a result, the exact physical mechanisms that govern the earliest stages of AGN activation remain elusive.

Mid-infrared emission provides a direct insight into obscured AGN activity, as dust heated by the central engine efficiently re-radiates absorbed ultraviolet and optical photons at infrared wavelengths \citep{2010A&A...509A..64K,2018ApJ...858...38S}. Multi-epoch mid-infrared observations represent a potentially powerful but unexplored approach in nascent AGN searches. Specifically, long-term infrared variability can reveal gradual changes in the nuclear energy output or the structure of the circumnuclear dust, even when optical variability is absent or suppressed by obscuration. A sustained, monotonic increase in mid-infrared emission over multi-year timescales is difficult to attribute to stellar processes, which typically evolve on much shorter timescales. Instead, such trends are naturally explained by a gradual increase in accretion power and a decrease in nuclear obscuration, as expected during the early stages of AGN turn-on.

NGC~6447 is an SB galaxy at $z=0.022462$ \citep{1999PASP..111..438F}, yielding a redshift distance of 98.4~Mpc. 
It has a Tully-Fisher distance of 78.4--84.5~Mpc \citep{2007A&A...465...71T}, presumably due to the relatively larger 21cm emission width of $284\pm48$~\kms \citep{1998A&AS..130..333T}.
The galaxy has an angular extent of 102\arcsec\ corresponding to 46.5~kpc.
NGC~6447 was considered a quiescent galaxy in the literature, and in particular, it was not classified as an AGN \citep{2019ApJ...872..134Z}.
In this paper, we present the decade-long monotonic brightening of the mid-infrared luminosity of NGC~6447, along with multiwavelength and timing data that resemble those of an AGN at the turning-on stage.
Throughout the paper, we assume cosmological parameters of H$_0$ = 70 km s$^{-1}$Mpc$^{-1}$, $\Omega_m$ = 0.3, and $\Omega_\Lambda$ = 0.7.

\section{Selection and Data Analysis}

NGC~6447 was selected from a Mid-IR variability-selected AGN sample from $\sim$7800 parent PanSTARRS galaxies with $g<14$~mag using the NEOWISE data.  The detailed selection will be presented in a subsequent paper.  The Mid-IR light curves of NGC~6447 shows unusual, almost monotonic increasing trend over a decade.  We subsequently analyzed archival data from ASAS-SN, WISE, SPHEREx, \nustar, and \swift\
from \citet{asassn_website}, \citet{IRSA2026},  \citet{heasarc_archive}
to comprehensively measure the timing and spectroscopic properties of the source.

The NEOWISE \citep{2014ApJ...792...30M} light curves in $W1$ (3.4$\mu$m) and $W2$ (4.6$\mu$m) bands were obtained by querying the
single-exposure, point-source database at the NASA/IPAC Infrared
Science Archive.  Since NEOWISE performed repeated scanning of the sky, we grouped the single-exposure observations into 180-day bins to reflect
this structure. Within each bin, outliers were rejected using a median absolute deviation
(MAD) filter, and the mean magnitude and sample standard deviation of the retained
measurements were calculated as the measurement and its uncertainty. To ensure reliable variability estimates, only bins containing at least three
single exposures were retained in the final sample.  We also included the WISE \citep{Wright2010} measurements in the MIR light curves.

ASAS-SN \citep{Shappee_2014, kochanek17} $V$ and $g$ band light curves were extracted using the image subtraction method from 2012--10--28 to 2025--06--15.
SPHEREx \citep{2025ApJ...985..100B} spectrum of NGC~6447 covers 0.75--4.2 $\mu$m at $R = 41$ and 4.2--5 $\mu$m at $R = 135$.  There are 130 total observations taken mainly from 2025--08--05 to 2025--08--24, with two on 2025--09--14 and 2025--10--02, by SPHEREx’s six detectors, each with an exposure time of 113.58~$s$.  Data with flags other than ``SOURCE'' are removed, resulting in 120 observations.

NGC~6447 was observed by \swift-XRT \citep{2004ApJ...611.1005G} and \nustar\ \citep{2013ApJ...770..103H} on 2023--08--27 and 2024--08--15 with exposures of 1538.3~$s$ and 40623.9~$s$, respectively.  
We downloaded the data from the HEASARC archive and analyzed the cleaned level 2 event files.
For the \nustar\ data, we used a circular source region with 0.82\arcmin\ radius and a background annulus with inner and outer radii of 2 and 2.8\arcmin, respectively, and analyzed FPMA and FPMB events together.
NGC~6447 is detected at 6.8$\sigma$, 3.8$\sigma$, and 7.5$\sigma$ above the background in the 3--10, 10--30, and 3--30~keV band, respectively, with count rates of $(6.6\pm1.0)\times10^{-3}$~ct~s$^{-1}$, $(4.7\pm1.2)\times10^{-3}$~ct~s$^{-1}$, $(11.3\pm1.5)\times10^{-3}$~ct~s$^{-1}$.
The unabsorbed flux of NGC~6447 is 2.7, 5.5, and 7.3$\times10^{-13}$~\flux\ in the 2--10, 10--30, and 2--30~keV band, using Galactic $\nh=2.27\times10^{20}$\cmsq\ and assuming a powerlaw index of $\Gamma=1.5$, yielding luminosities of 3.1, 6.2, and 8.4$\times10^{41}$~\lumin\ in the three bands.
The \swift-XRT observation was made in the photon counting mode, and we used a circular source region with 18\arcsec\ radius and a background annulus with inner and outer radii of 2\arcmin\ and 4\arcmin\ to extract source and background photons, yielding zero source photons in the source region.  We estimated a 3$\sigma$ upper limit of $1.2\times10^{-3}$~ct~s$^{-1}$ corresponding to a 2--10~keV flux limit of $3.7\times10^{-14}$\flux, assuming the same spectral model for the \nustar\ flux conversion.


\begin{figure*}[ht!]
\epsscale{1}
\plotone{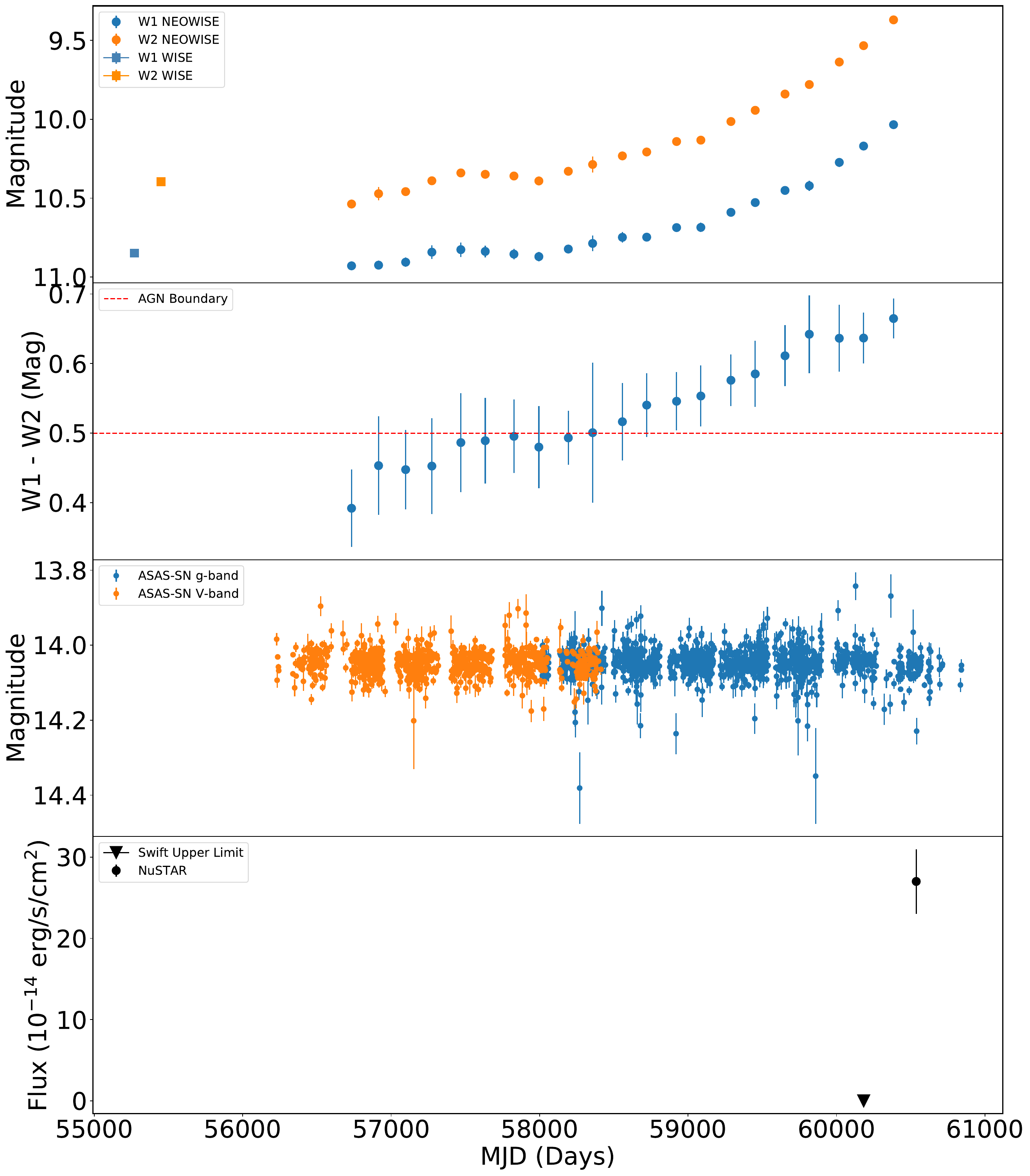}
	\caption{Light curves of NGC~6447 in MIR, optical, and X-ray bands. The top panel shows the WISE/NEOWISE W1 and W2 light curves over 15 years, where an almost monotonic brightening trend is shown in both W1 and W2 light curves.  Second panel: the W1$-$W2 color evolution, where the W1$-$W2 color moved into the AGN regime in 2018 based on the classifications of \citet{2018MNRAS.478.3056B}.  Third panel: ASAS-SN optical light curve in V and g bands, where V and g magnitudes are normalized. The optical light curve is consistent with a constant.  Bottom Panel: 2--10~keV X-ray light curve from 3$\sigma$ \swift\ upper limit and \nustar\ detections.  Significant variability by a factor $>7.3$ is detected in the X-ray band.  The 2-10~keV X-ray luminosity is $3.1\times10^{41}$~\lumin.
	\label{fig:lc}}
\end{figure*}

\begin{figure*}[ht!]
\epsscale{0.7}
\plotone{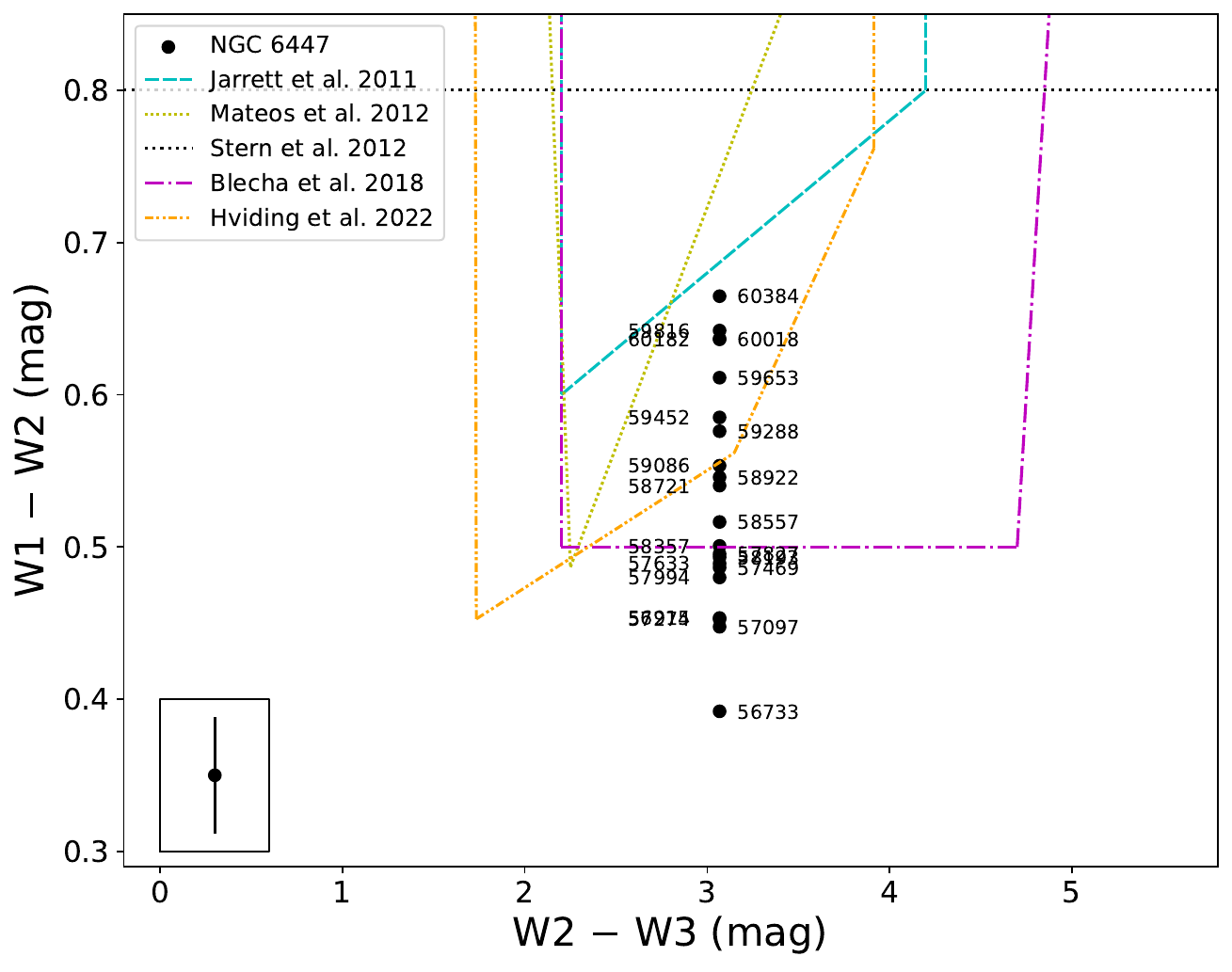}
	\caption{MIR color evolution of NGC~6447 based on the NEOWISE light curves, which evolves deeper into the AGN region specified by \citet{2018MNRAS.478.3056B, 2022AJ....163..224H}. 
    Other MIR AGN selection criteria plotted are from \citet{2011ApJ...735..112J, 2012ApJ...753...30S, 2012MNRAS.426.3271M}.  We assume that the W2$-$W3 color is the same as measured by WISE. A typical uncertainty is plotted at the bottom-left corner.
    \label{fig:color}}
\end{figure*}

\begin{figure*}[ht!]
	\plotone{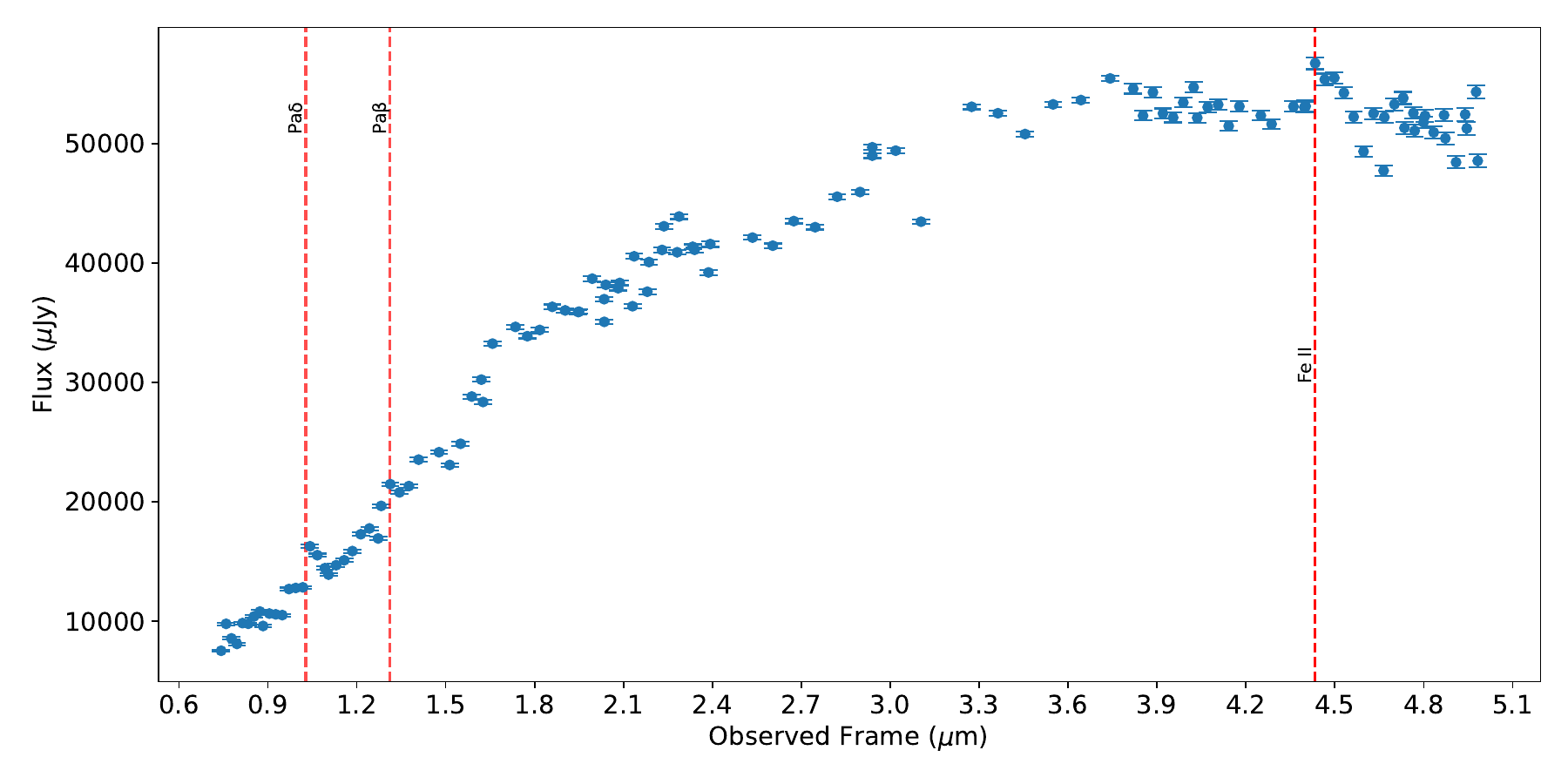}
	\caption{SPHEREx spectrum of NGC~6447, showing characteristics of AGN emission. The rising continuum represents warm and hot dust emission from the central engine.  A few excess emission features are identified as \FeII, Pa$\beta$, and Pa$\delta$ lines. \label{fig:spec}}
\end{figure*}

\section{Results and Discussion} \label{sec:res}
We show the light curves of NGC~6447 in MIR, optical, and X-rays in Figure~\ref{fig:lc}.
The NEOWISE MIR light curves in both the W1 and W2 bands show a distinct, almost monotonic brightening over a decade from 2014--03 to 2024--03 by 0.9 and 1.2~mag, respectively.
The WISE photometry in W1 and W2 bands taken in 2010 is largely consistent with the NEOWISE light curve trend (Figure~\ref{fig:lc}).
The optical ASAS-SN light curve from 2012--11 to 2025--06 exhibits no variability with an excess variable value consistent with values of non-variable sources \citep{2025ApJ...985..177K} from the ASAS-SN survey.
The long rising MIR light curve, coupled with no optical variability, suggests that the optical emitting disk is still hidden by dust obscuration, and MIR emission of the central engine is gradually revealed as the central engine disperses dust away.
The W1$-$W2 color evolution shows that NGC~6447 transitioned from a galaxy to AGN around 2018--08, based on the MIR AGN color selection criterion of \citet{2018MNRAS.478.3056B}.
Before the transition epoch, W1 and W2 light curves brightened by 0.15 and 0.25~mag, and after the transition, the light curves brightened by 0.75 and 0.91~mag, showing an increasing brightening trend.
Two X-ray measurements of NGC~6447 are available, and they show significant variability by a factor of $>7.3$ within one year in the 2--10~keV band, and the \nustar\ detection measures a luminosity of 8.4$\times10^{41}$\lumin\ in the 2--30~keV band. 
Both the X-ray variability and luminosity point to an AGN in NGC~6447.
The MIR color-color diagram (Figure~\ref{fig:color}) better illustrates the transitioning of NGC~6447, as the source originally classified as non-AGN gradually enters various AGN classification boundaries \citep{2018MNRAS.478.3056B, 2022AJ....163..224H}.  Here, the W3 magnitude was measured by WISE, and we assume the W2$-$W3 color remains the same during the period.
SPHEREx spectrum of NGC~6447 covers from 0.6 to 5.1$\mu$m, showing a rising continuum in MIR (Figure~\ref{fig:spec}), resembling warm to hot dust emission in AGN central  \citep{1989ApJ...347...29S, 2010A&A...509A..64K, 2018ApJ...858...38S}.  We identify two emission features with \FeII\ and Pa$\delta$ lines, and Pa$\beta$ is potentially associated with a low S/N emission feature. These emission features are consistent with emission lines from the broad line regions \citep{2008ApJS..174..282L}.
The multi-wavelength timing and spectroscopic data are consistent with a picture that NGC~6447 is at the turning-on stage of galaxy evolution, where the central engine is in the process of dispersing the obscuration dust and reveal the AGN features in the MIR and X-ray bands and the still remaining obscuration keeps the optical disk emission hidden \citep{1995Natur.375..469W, 2004ApJ...600..580G, 2007ApJ...667..673G, 2012ApJ...757...51G}.

Beside the AGN central engine, a supernova can produce variability with the amplitude matching the host galaxy.  
However, the timescale of supernova explosion, a few months, especially the rising timescale is much shorter, with only tens of days \citep{Leibundgut2000, Goobar2011}, which does not match the observation of NGC~6447. Tidal disruption events (TDE), in which a star is torn apart by a supermassive black hole, are also unlikely to explain the observed behavior. TDEs produce rapid optical/ultraviolet flares over timescales of days or weeks, followed by a decline over timescales of months to a few years \citep{1988Natur.333..523R, 2017ApJ...842...29H}. The flares are often accompanied by mid-infrared dust echoes that rise on timescales of a few years at most following the initial outburst as circumnuclear dust reprocesses the transient emission \citep{2021ApJ...911...31J}. In contrast, in NGC6447, no significant optical variability is detected, whereas infrared emission exhibits a decade-long, spectrally evolving monotonic rise. Additionally, TDE X-ray emission is usually detected near the peak of the event, often at high luminosities \citep{2021ApJ...908....4V}. Therefore, a delayed detection of relatively low-luminosity X-ray emission is inconsistent with standard TDE evolution.

A persistent low-luminosity AGN (LLAGN) is also unlikely. LLAGN, typically powered by radiatively inefficient accretion flows, usually have low Eddington ratios, relatively stable luminosities, and rarely exhibit short-term, large-amplitude X-ray variability \citep{2008ARA&A..46..475H,1998ApJ...501L..37P}. The rapid increase of NGC 6447 X-ray emission by a factor of $>$7.3 is difficult to describe with a steady LLAGN and instead signals the onset of a new radiatively efficient accretion episode. Additionally, a recent study of a sample of optical variability-selected LLAGN shows very little overlap with MIR color-selected AGN \citep{2025ApJ...987..112Y}.

Changing-look AGN usually refers to luminous AGN with dramatic spectral changes between broad and narrow line AGN \citep{2003MNRAS.342..422M}, blazar types \citep{2021ApJ...913..146M}, or significant spectral line variability \citep{2024ApJ...966...85Z}. The turning-on case of NGC~6447 can be considered as a special changing-look AGN, but with significant characteristics differences from other changing-look AGN mentioned above.  Future multi-wavelength timing and spectroscopic observations will be important to further confirm the nature of NGC~6447 evolution.

In the remaining discussion, we assume that a turning-on AGN is the best interpretation of the timing and spectroscopic data of NGC~6447.
We can estimate the rate of such events based on the sample selection statistics, where only one such example is found in $\sim$7800 galaxies, or a rate of 0.013\% among galaxies.  Compared to the AGN fraction from a few to $\sim10$\%, about 0.13\% to 0.26\% of AGN are on this stage. Assuming a global AGN growth timescale of $\sim10^{7}$--$10^{9}$~yr based on Soltan argument \citep{1982MNRAS.200..115S, 2002MNRAS.335..965Y, 2004MNRAS.351..169M}, each episode of growth evidenced by the turning-on stage would last $\sim10^{4}$--$10^{6}$~yr, by applying the above fractions of 0.13\% to 0.26\%.

This is broadly consistent with the general model predictions \citep{2011ApJ...737...26N, 2013MNRAS.434..606G}. Observationally, the AGN growth episodes can be constrained using the turning-off AGN, identified from the extended emission line regions.
\citet{2015MNRAS.451.2517S} estimated the individual AGN ``flicker'' phase of $\sim10^{5}$~yr based on the lag between AGN X-ray variability and host photoionization properties.
Using radiative transfer simulations, a more recent work estimated the AGN duty cycle as $10^{4}$--$10^{6}$~yr \citep{2025A&A...703A..63F}.
From the IGM properties in the proximity zone around quasars, which encodes quasar's ionization history, the duty cycle of quasars are estimated to be $\sim10^{6}$--$10^7$~yr \citep{2016ApJ...824..133K, 2021ApJ...917...38E}.

The example of NGC~6447 shows that long-term IR light curves are powerful diagnostics to search for AGN at special evolutionary stages.  A more dedicated effort to search for these cases from a much larger sample is under investigation, where we expect to better constrain the occurrence statistics of these turning on AGN.  As the early stage of turning-on candidates is extremely rare, detailed follow-up studies of NGC~6447 will better constrain the details of the physical process and theoretical models.

\begin{acknowledgments}
We would like to acknowledge NASA funds 80NSSC22K0488, 80NSSC23K0379, and NSF fund AAG2307802. 
\end{acknowledgments}

\facilities{NEOWISE, ASAS-SN, SPHEREx, NuSTAR, Swift}




\bibliography{ngc6447}{}
\bibliographystyle{aasjournalv7}

\end{document}